\documentstyle[12pt]{article}

\catcode`\@=11
\def\@citex[#1]#2{%
\if@filesw \immediate \write \@auxout {\string \citation {#2}}\fi
\@tempcntb\m@ne \let\@h@ld\relax \def\@citea{}%
\@cite{%
  \@for \@citeb:=#2\do {%
    \@ifundefined {b@\@citeb}%
      {\@h@ld\@citea\@tempcntb\m@ne{\bf ?}%
      \@warning {Citation `\@citeb ' on page \thepage \space undefined}}%
      {\@tempcnta\@tempcntb \advance\@tempcnta\@ne%
      \@tempcntb\number\csname b@\@citeb \endcsname \relax%
      \ifnum\@tempcnta=\@tempcntb %   Number follows previous--hold on to it
        \ifx\@h@ld\relax%
          \edef \@h@ld{\@citea\csname b@\@citeb\endcsname}%
        \else%
          \edef\@h@ld{\ifmmode{-}\else--\fi\csname b@\@citeb\endcsname}%
        \fi%
      \else%   %  non-successor--dump what's held and do this one
        \@h@ld\@citea\csname b@\@citeb \endcsname%
        \let\@h@ld\relax%
      \fi}%
    \def\@citea{,\penalty\@highpenalty\,}%
  }\@h@ld
}{#1}}
\newcount\hour
\newcount\minute
\newtoks\amorpm
\hour=\time\divide\hour by 60
\minute=\time{\multiply\hour by 60 \global\advance\minute by-\hour}
\edef\standardtime{{\ifnum\hour<12 \global\amorpm={am}%
        \else\global\amorpm={pm}\advance\hour by-12 \fi
        \ifnum\hour=0 \hour=12 \fi
        \number\hour:\ifnum\minute<10 0\fi\number\minute\the\amorpm}}
\edef\militarytime{\number\hour:\ifnum\minute<10 0\fi\number\minute}

\def\draftlabel#1{{\@bsphack\if@filesw {\let\thepage\relax
   \xdef\@gtempa{\write\@auxout{\string
      \newlabel{#1}{{\@currentlabel}{\thepage}}}}}\@gtempa
   \if@nobreak \ifvmode\nobreak\fi\fi\fi\@esphack}
        \gdef\@eqnlabel{#1}}
\def\@eqnlabel{}
\def\@vacuum{}
\def\marginnote#1{}
\def\draftmarginnote#1{\marginpar{\raggedright\scriptsize\tt#1}}

\def\draft{\oddsidemargin -.5truein
        \def\@oddhead{\sl \phantom{\today\quad\militarytime} \hfil
        \smash{\Large\sl DRAFT} \hfil \today\quad\militarytime}
        \let\@evenhead\@oddhead
        \let\label=\draftlabel
        \let\marginnote=\draftmarginnote
        \def\ps@empty{\let\@mkboth\@gobbletwo
        \def\@oddfoot{\hfil \smash{\Large\sl DRAFT} \hfil}
        \let\@evenfoot\@oddhead}
        \def\@eqnnum{(\theequation)\rlap{\kern\marginparsep\tt\@eqnlabel}%
        \global\let\@eqnlabel\@vacuum}  }

\def\nblack{            % For people without blackboard fonts
\def\ZZ{Z \n{10} Z}
\def\NN{N \n{14} N}
\def\CC{C \n{11} C}
\def\RR{R \n{11} R}
\def\QQ{Q \n{12} Q}
\def\PP{P \n{11} P}
}

\def\section{\@startsection {section}{1}{\z@}{3.ex plus 1ex minus
 .2ex}{2.ex plus .2ex}{\large\bf}}
\def\subsection{\@startsection{subsection}{2}{\z@}{2.75ex plus 1ex minus
 .2ex}{1.5ex plus .2ex}{\bf}}

\def\appendix{\let\appendix\moreappendix%
        \setcounter{section}{0} \setcounter{subsection}{0}
        \@addtoreset{theorem}{section}
        \def\thesection{\Alph{section}}\moreappendix}
\def\moreappendix{\@startsection {section}{1}{\z@}{3.ex plus 1ex minus
 .2ex}{2.ex plus .2ex}{\large\bf Appendix }}

\def\abstract{\if@twocolumn
\section*{Abstract}
\else %\small
\begin{center}
{\bf Abstract\vspace{-.5em}\vspace{0pt}}
\end{center}
\quotation
\fi}

\def\eqalign#1{\null\,\vcenter{\openup\jot\m@th
  \ialign{\strut\hfil$\displaystyle{##}$&$\displaystyle{{}##}$\hfil
      \crcr#1\crcr}}\,}
\def\eqalignno#1{\displ@y \tabskip\centering
  \halign to\displaywidth{\hfil$\@lign\displaystyle{##}$\tabskip\z@skip
    &$\@lign\displaystyle{{}##}$\hfil\tabskip\centering
    &\llap{$\@lign##$}\tabskip\z@skip\crcr
    #1\crcr}}

\catcode`\@=12

\def\eq#1{.~(\ref{#1})}
\def\beq{\begin{equation}}
\def\eeq{\end{equation}}
\def\beqar{\begin{eqnarray}}
\def\eeqar{\end{eqnarray}}

\def\nfrac#1#2{{\displaystyle{\vphantom1\smash{\lower.5ex\hbox{\small$#1$}}%
        \over\vphantom1\smash{\raise.25ex\hbox{\small$#2$}}}}}

\def\p#1{\mskip#1mu}
\def\n#1{\mskip-#1mu}
\def\stop{\p6.}
\def\comma{\p6,}

\def\noj#1,#2,{{\bf #1} (19#2)\ }
\def\jou#1,#2,#3,{{\sl #1\/ }{\bf #2} (19#3)\ }
\def\ann#1,#2,{{\sl Ann.\ Physics\/ }{\bf #1} (19#2)\ }
\def\cmp#1,#2,{{\sl Comm.\ Math.\ Phys.\/ }{\bf #1} (19#2)\ }
\def\cq#1,#2,{{\sl Class.\ Quantum Grav.\/ }{\bf #1} (19#2)\ }
\def\cqg#1,#2,{{\sl Class.\ Quantum Grav.\/ }{\bf #1} (19#2)\ }
\def\ijmp#1,#2,{{\sl Int.\ J.\ Mod.\ Phys.\/ }{\bf A#1} (19#2)\ }
\def\jmp#1,#2,{{\sl J.\ Math.\ Phys.\/ }{\bf #1} (19#2)\ }
\def\grg#1,#2,{{\sl Gen.\ Rel.\ Grav.\/ }{\bf #1} (19#2)\ }
\def\mpl#1,#2,{{\sl Mod.\ Phys.\ Lett.\/ }{\bf A#1} (19#2)\ }
\def\nc#1,#2,{{\sl Nuovo Cim.\/ }{\bf #1} (19#2)\ }
\def\np#1,#2,{{\sl Nucl.\ Phys.\/ }{\bf B#1} (19#2)\ }
\def\pl#1,#2,{{\sl Phys.\ Lett.\/ }{\bf #1B} (19#2)\ }
\def\pla#1,#2,{{\sl Phys.\ Lett.\/ }{\bf #1A} (19#2)\ }
\def\pr#1,#2,{{\sl Phys.\ Rev.\/ }{\bf #1} (19#2)\ }
\def\prd#1,#2,{{\sl Phys.\ Rev.\/ }{\bf D#1} (19#2)\ }
\def\prl#1,#2,{{\sl Phys.\ Rev.\ Lett.\/ }{\bf #1} (19#2)\ }
\def\prp#1,#2,{{\sl Phys.\ Rept.\/ }{\bf #1C} (19#2)\ }
\def\ptp#1,#2,{{\sl Prog.\ Theor.\ Phys.\/ }{\bf #1} (19#2)\ }
\def\ptpsup#1,#2,{{\sl Prog.\ Theor.\ Phys.\/ Suppl.\/ }{\bf #1} (19#2)\ }
\def\rmp#1,#2,{{\sl Rev.\ Mod.\ Phys.\/ }{\bf #1} (19#2)\ }
\def\yadfiz#1,#2,#3[#4,#5]{{\sl Yad.\ Fiz.\/ }{\bf #1} (19#2) #3%
\ [{\sl Sov.\ J.\ Nucl.\ Phys.\/ }{\bf #4} (19#2) #5]}
\def\zh#1,#2,#3[#4,#5]{{\sl Zh.\ Exp.\ Theor.\ Fiz.\/ }{\bf #1} (19#2) #3%
\ [{\sl Sov.\ Phys.\ JETP\/ }{\bf #4} (19#2) #5]}

\def\adhoc{{\it ad hoc\/}}

\def\ansatz{{\it ansatz\/}}
\def\ansatze{{\it ans\"atze\/}}

\def\mobius{m\"obius}
\def\Mobius{M\"obius}

\def\etal{{\it et al.\/}}
\def\etc{{\it etc.\/}}

\def\to{\rightarrow}
\def\implies{\Rightarrow}

\def\longlongrightarrow{\relbar\joinrel\relbar\joinrel\rightarrow}

\def\onArrow#1{\mathrel{\mathop{\longlongrightarrow}\limits^{#1}}}

\def\lae{\mathrel{\mathop{\smash{\lower .5 ex \hbox{$\stackrel<\sim$}}}}}
\def\lae{\mathrel{\mathop{\smash{\lower .5 ex \hbox{$\stackrel>\sim$}}}}}
\def\eqq{\stackrel?=}

\def\vev#1{\left\langle #1 \right\rangle}
\def\VEV#1{\left\langle #1 \right\rangle}
\def\f{\frac}
\def\pa{\partial}
\def\pb{\bar\pa}

\def\Tr{{\rm Tr}}
\def\l:{\mathopen{:}\,}
\def\r:{\,\mathclose{:}}

\def\baselinestretch{1.2}
\nblack

\catcode`\@=11
\@addtoreset{footnote}{section}
\@addtoreset{footnote}{subsection}
\catcode`\@=12

\def\Kahler{K\"ahler}
\def\nq{$N=2$}
\def\twot{$(2,2)$}
\def\xb{\bar{x}}
\def\jb{\bar{j}}
\def\ib{\bar{i}}
\def\wb{\bar{w}}
\def\zb{\bar{z}}
\def\Xb{\bar{X}}
\def\kb{\bar{k}}
\def\pb{\bar{\pa}}
\def\tb{\bar{\theta}}
\def\psb{\bar{\psi}}

\def\dx{d x}
\def\dxf{d^{\p1 4}\n2 x}
\def\dtt{d^{\p1 2}\theta}

\def\sut{SU(2)}

\def\son{SO(N)}
\def\sot{SO(2)}
\def\spn{USp(N)}
\def\sott{SO(2,2)}
\def\uo{U(1)}
\def\uoo{U(1,1)}

\def\sltrlr{\sltr_L\otimes\sltr_R}
\def\sltr{SL(2,\RR)}
\def\rptwo{RP_2}

\def\klein{\hbox{\scriptsize\sl Klein}}
\def\annulus{\hbox{\scriptsize\sl annulus}}
\def\torus{\hbox{\scriptsize\sl torus}}
\def\mobius{\hbox{\scriptsize\sl M\"obius}}

\begin{document}
\begin{titlepage}

\begin{center}
July 9, 1992\hfill    TAUP--1929--91 \\
\hfill    hep-th/yymmddd

\vskip 1 cm

{\large \bf
The $N=2$ open string.
}

\vskip 1 cm

{
          Neil Marcus\footnote{
Work supported in part by the US-Israel Binational Science Foundation,
and the Israel Academy of Science.\\E-Mail:
NEIL@HALO.TAU.AC.IL.}
}

\vskip 0.2cm

{\sl
School of Physics and Astronomy\\Raymond and Beverly Sackler Faculty
of Exact Sciences\\Tel-Aviv University\\Ramat Aviv, Tel-Aviv 69978, ISRAEL.
}

\end{center}

\vskip 1 cm

\begin{abstract}

We show that the $N=2$ open string describes a theory of self-dual Yang
Mills (SDYM) in \twot{} dimensions.  The coupling to the closed sector is
described by SDYM in a \Kahler{} background, with the Yang-Mills fields
providing a source term to the self-duality equation in the gravity sector.
The four-point S-matrix elements of the theory vanish, so the tree-level
unitarity constraints leading to the Chan-Paton construction are relaxed.
By considering more general group-theory \ansatze{} the $N=2$ string can be
written for any gauge group, and not just the classical groups allowed for
the bosonic and $N=1$ strings.  Such \adhoc{} group-theory factors
can not be appended to the closed $N=2$ string, explaining why the
$\ZZ_n$ closed $N=2$ strings are trivial extensions of the $\ZZ_1$ theory.

\end{abstract}
\newpage

\end{titlepage}

\section{Introduction.}

\nq{} strings have had a short but rather convoluted history.  The \nq{} open
string was first studied by M.~Ademollo \etal{} \cite{adem}.
They found that the critical dimension of the theory was $D=2$ and
that the theory contained only massless scalars in the adjoint of the
\sut{} Chan-Paton group.  The on-shell three-point vertex vanished,
but the theory had nontrivial local four-point vertices.  It was argued that
only $4N$-point vertices existed, and that the theory could
be described by an effective sigma-model field theory.

Many years later, M.B.~Green considered also closed and heterotic \nq{}
strings in two dimensions \cite{Mike}.  These heterotic strings are
classified by the 24-dimensional lattice of the bosonic sector, which
could be, for instance, an $E_8^3$ lattice or a Moonshine module.  While
such heterotic strings have interactions similar to those of
the open theory, the closed string was seen to have a trivial
S-matrix.  The effective theory of the heterotic string was again
found to be a sigma model, but with corrections coming from diagrams
with intermediate gravitons.

At about this time it was pointed out by A.~D'Adda and F.~Lizzi
that, because of the \uo{} symmetry structure of the \nq{} Virasoro algebra
\cite{ademonly}, the theory should be considered to exist in two
{\it complex,\/} or four real, dimensions \cite{4d}.
They therefore argued that
the theory should have an $\sott{} \sim \sltrlr$ spacetime Lorentz
group\footnote{We
shall always consider the Minkowski space theories.}  and,
since the four-point function vanished, argued that
the open theory in four dimensions had a trivial S matrix.

Recently, H.~Ooguri and C.~Vafa re-examined the closed \nq{} string
\cite{OV}.  They took the theory to live in an intrinsically \Kahler{}
two-complex dimensional spacetime, with Lorentz group $\uoo \sim \sltr
\otimes \uo$.  They found that the four-point S matrix elements vanish
also for this case, but noted that the {\it three-point\/} function
does not vanish on shell, so the theory has a nontrivial S matrix.  They
argued that all higher-point functions should vanish, and interpreted
the theory as having a topological or integrable structure.  The
massless scalar of the theory was interpreted as the \Kahler{}
potential of the spacetime.  It satisfies a quadratic equation of motion, the
Plebanski equation \cite{Plebanski}, which is the condition that the
spacetime have a self-dual Riemann tensor.  Thus the closed \nq{} string can
be thought of as a theory of self-dual gravity in \twot{} dimensions.  A
$\ZZ_n$ generalization of this theory, containing $n$ particles with
spins under the $\uo$ part of the Lorentz group, was also studied in
ref.~\cite{OV}.

Ooguri and Vafa also reconsidered the \nq{} heterotic string in the
light of these new developments \cite{OVh}.  Since the left-handed
\nq{} string has to be married to a bosonic or superstring with only
one time direction, it turns out that the dependence of the theory on a
timelike (and sometimes also a spatial) direction has to be fixed,
resulting in a theory in a $(2,1)$ or a $(1,1)$ dimensional subspace. The
equation of motion in the gauge sector of the three-dimensional theory
\cite{OVh} is essentially a dimensional reduction of Yang's equation for
self-dual Yang Mills (SDYM) in four dimensions \cite{Yang} (which can be
obtained from the five-dimensional action of Nair and Schiff \cite{NS}).
Somewhat surprisingly, the resulting three-dimensional scalars of the
theory are tachyonic! In addition, the theory also contains massless
vector-like particles in the gravitational sector, whose couplings to the
scalars are somewhat messy and poorly understood.  Diagrams with
intermediate vector particles induce an $O(\alpha')$ modification to the
equation of motion of the scalars, so that the gauge sector of the theory
is not simply SDYM. The two-dimensional theory is also a dimensional
reduction of Yang's equation, which is equivalent to
the equation of motion of the sigma model.  Ref.~\cite{Mike}
shows that the theory has $O(\alpha')$ corrections modifying the SDYM (or
sigma model) structure also in this case, in disagreement with
ref.~\cite{OVh}.

In this paper, we shall reconsider the \nq{} open string, now taken to be
in \twot{} dimensional spacetime.  In section 2 we calculate
several three and four point tree-level amplitudes of open and closed strings.
In section~3 we consider the effective field theory of the open
string, and in section~4 we give our conclusions.

\section{\nq{} open-string amplitudes.}

The tree-level action of open strings is found both
from genus zero amplitudes of the closed
sector and ``genus $1/2$'' amplitudes of the open and closed sectors.
In order to find the genus zero amplitudes, to
illustrate the issues involved in working in a \twot{} space, and to
establish notation, we shall, with apologies, first briefly review
section~2 of ref.~\cite{OV}.

\subsection{The closed string and its amplitudes.}

The superspace-vertex operator for
emitting a closed-string scalar of momentum $k$ is\footnote{Here the
``dots'' indicate contractions over the two complex coordinates.}
\beq
V_c = \f\kappa\pi \, e^{i(k \cdot \Xb + \kb \cdot X )} \comma
\eeq
where $X$ is an \nq{} chiral superfield.  Thus, the vertex at
$\theta=\tb=0$ is simply the momentum insertion
\beq
V_c^0 |_{\theta=\tb=0} = \f\kappa\pi \, e^{i(k \cdot \xb + \kb \cdot x )}
\comma
\label{vctriv}
\eeq
while the vertex integrated over the fermionic coordinates is
\beq
V_c^{int} = %\int \dtt \, \dttb \, V_c =
\f\kappa\pi \,
( i k \cdot \pa \xb -i \kb \cdot \pa x - k \cdot \psb_R \, \kb \cdot \psi_R )
\,( i k \cdot \pb \xb -i \kb \cdot \pb x - k \cdot \psb_L \, \kb \cdot \psi_L )
\, e^{i(k \cdot \xb + \kb \cdot x )} \stop
\label{vcmess}
\eeq

To obtain the three-point function, one needs to fix three bosonic
coordinates and two $\theta$'s (and $\tb$'s).
Thus one calculates the correlation function of
one $V_c^{int}$ and two $V_c^0$'s, at fixed positions.  The result
is\footnote{We differ from ref.~\cite{OV} by a factor of $\pi$.  See,
for example, ref.~\cite{openclosed}.}
\beq
A_{ccc} = \kappa \, c_{12}^2 \comma
\label{a3c}
\eeq
where
\beq
c_{12} \equiv ( k_1 \cdot \kb_2 -\kb_1 \cdot k_2 )
\eeq
is the extra invariant product of the momenta (other than the dot
product) that exists when the
Lorentz group is reduced to $\sltr \otimes \uo$.  Note that $c_{ij}$
is antisymmetric with respect to its two indices, and is additive in
the sense that $c_{i,j}+c_{i,k} = c_{i,j+k}$.  Using momentum
conservation, one sees that $A_{ccc}$ is totally symmetric, as it should
be.  The four-point function is given by the
correlation function of two $V_c^{int}$'s and two $V_c^0$'s,
with the position of
one of the vertices integrated over the sphere.
The result is
\beq
A_{cccc} = \f{\kappa^2}{\pi}
         F^2 \, \f { \Gamma(1-s/2) \, \Gamma(1-t/2) \, \Gamma(1-u/2) }
                 { \Gamma(s/2) \, \Gamma(t/2) \, \Gamma(u/2) } \stop
\label{a4c}
\eeq
The crucial point of ref.~\cite{OV} is that the kinematic factor
\beq
F \equiv 1 - \f{c_{12}c_{34}}{su} - \f{c_{23}c_{41}}{tu}
\label{F}
\eeq
vanishes on shell in the scattering of massless particles in \twot{}
dimensions, so the four-point function of the
theory vanishes identically.  The local three-point function and
vanishing four-point function can then be obtained from
the action \cite{OV}
\beq
{\cal L}_c = \int \dxf \, \left( \nfrac12 \, \pa^i \phi \, \pb_i \phi +
\f{2\kappa}3 \, \phi \, \pa \pb \phi \wedge \pa \pb \phi \right) \stop
\label{pleb}
\eeq
The equation of motion of this action is the
Plebanski equation \cite{Plebanski}---the equation for self-dual
gravity written in terms of the \Kahler{} potential in the \twot{} space.
In ref.~\cite{OV}, it is argued that this cubic action is exact.

Note that if the theory is restricted to two {\it real} dimensions, the
invariant $c_{12}$ vanishes and the theory has no three-point vertex.  The
kinematic factor $F$ is now 1, but the four-point function of eq\eq{a4c}
still vanishes,
since $stu=0$ in the scattering of massless particles in two dimensions.

\subsection{Open string amplitudes and group theory.}

As argued in ref.~\cite{adem} in two real dimensions, and in
ref.~\cite{4d} in two complex dimensions, the only particle in the
open sector of the \nq{} theory is a massless scalar (which will
become a multiplet of scalars after the addition of  group-theory factors to
the theory).
The superspace vertex for emitting the scalar is
\beq
V_o = g \, e^{i(k \cdot \Xb + \kb \cdot X )} \comma
\eeq
where the vertex now lives on the boundary of the super-Riemann
surface: $z = \zb \equiv \sigma$, $\theta = \tb \equiv \theta$.
The boundary
conditions of the fields are $\pa x = \pb x |_{z=\zb}$ and $\psi_R=\psi_L
\equiv \psi |_{z=\zb}$.
Integrating out the fermionic coordinates, one obtains\footnote{The strange
factors of
2 are the result of using closed-string conventions.}
\beq
V_o^{int} = \int \dtt \, V_o =
\f{g}2 \, ( i k \cdot \pa_\sigma \xb -i \kb \cdot \pa_\sigma x -
4 k \cdot \psb \, \kb \cdot \psi )
\, e^{i(k \cdot \xb + \kb \cdot x )} \stop
\label{vomess}
\eeq

The three-point open-string amplitude is again found from the correlation
function of one
$V_o^{int}$ and two $V_o^0$'s at fixed positions.
As usual \cite{openclosed}, the result is basically the square root
of the closed-string amplitude of eq\eq{a3c}:
\beq
A_{ooo} = g \, c_{12} \stop
\eeq
This amplitude is totally antisymmetric with respect to
the three scalars, so one
needs to insert group-theory factors to prevent
the vanishing of the amplitude.  We shall not restrict ourselves to
the standard Chan-Paton factors \cite{CP}, but shall rather take the
most general possible \ansatz{} consistent with principles to be given
later.  We thus add an index $a$ to the open-string scalars,
taking their kinetic term to be
diagonal with a group factor of $\delta^{ab}$.  The most general
three-point function is now
\beq
A_{ooo} = -i g \, c_{12} f^{abc} \comma
\label{a3o}
\eeq
with
$f^{abc}$ an as yet unspecified totally antisymmetric tensor.
(In the Chan-Paton \ansatz, one would have $f^{abc}=
\Tr \, ( \Lambda^a [ \Lambda^b \Lambda^c ] ) $.)
Unsurprisingly, the vertex of eq\eq{a3o} is the same as that of the heterotic
string, where the $f^{abc}$'s are the structure constants of the SDYM
gauge group \cite{OVh}.  However, unlike the heterotic case,
the theory is defined in the full \twot{} dimensional space,
and does not need to be dimensionally reduced.

Temporarily putting aside the issue of the group theory, we now
proceed to the four-point function.  This amplitude
depends on the cyclic ordering of the vertices on the boundary, which
we choose to place at 0, $x$, 1 and $\infty$, with
$x$ integrated between 0 and 1.  Two of the vertices are
of the type $V_o^{int}$, and the other two are $V_o^0$'s.
The result is\footnote{This
amplitude was essentially obtained in ref.~\cite{Corvi}, but there it was
not realized that it vanishes on shell, giving a very
different interpretation of the result. }
\beq
\eqalign{
A_{oooo} &= g^2
            \int_0^1 \dx \left( \f{t (t+1/2)}{(1-x)^2} +\,  \f{c_{12} c_{34}}z
+\,  \f{c_{23}c_{41}}{1-x} \right) \, x^{-2s} (1-x)^{-2t} \cr
&= \f{g^2}{4} \,
F \; \f { \Gamma(1-2s) \, \Gamma(1-2t)} {\Gamma(2u) } \comma
\label{a4o}
}
\eeq
with $F$ the vanishing kinematic factor of eq\eq{F}.  Thus the
four-point amplitude is zero, and the theory is ``topological''
in the sense of Ooguri and Vafa.  Note that if the theory is reduced
to two real dimensions the three-point function vanishes, but
the four-point function becomes
\beq
A_{oooo} \onArrow{D \to 2} \, \f{g^2}2 \,u \comma
\eeq
in agreement with the result of Ademollo \etal{} \cite{adem}.

Since the four-point function vanishes, the usual factorization constraints
\cite{meold} leading to the Chan-Paton \ansatz{} \cite{CP}
no longer apply.  However,
one does find a constraint on the $f^{abc}$'s by demanding consistency
between the vanishing four-point amplitude of the string theory and
the same amplitude calculated from the effective field theory.  The
three-point amplitude of eq\eq{a3o} can be obtained from the
lagrangian \cite{OVh}
\beq
{\cal L}_{3o} = \int \dxf \, \left( \nfrac12 \, \pa^i \varphi^a \pb_{\ib}
\varphi^a -
i \,\f{g}3 \, f^{abc} \varphi^a \pa^i \varphi^b \pb_{\ib} \varphi^c \right)
\stop
\label{cubic}
\eeq
The field-theory four-point function is now calculated by sewing together the
vertices (\ref{a3o}) from this action, and adding some local
four-point vertex $V_{4o}$.  This gives
\beq
A_{oooo\,\hbox{\scriptsize FT}} = -g^2 \left\{ A f^{abx} f^{xcd} +
B f^{bcx} f^{xda} + C f^{cax} f^{xbd} \right\} - V_{4o} \comma
\eeq
where we have defined
\beq
A = \f{c_{12}c_{34}}s \qquad
B = \f{c_{23}c_{41}}t \qquad
C = \f{c_{31}c_{24}}u
\eeq
for the poles of the four-point amplitude in the three channels.  The
kinematic relation $F=0$ and its permutations imply that
\beq
B = u - A \qquad \hbox{and} \qquad C = t+A  \stop
\label{iden}
\eeq
These identities allow us to reduce the amplitude to
\beq
A_{oooo\,\hbox{\scriptsize FT}}
= -g^2 \left\{
A \left ( f^{abx} f^{xcd} - f^{acx} f^{xbd} - f^{bcx} f^{xda} \right )
+ u f^{bcx} f^{xda} + t f^{cax} f^{xbd} \right\} - V_{4o} \stop
\label{V4}
\eeq
For the amplitude to be zero, in agreement with eq\eq{a4o}, the factor
multiplying $A$ must vanish, since $V_{4o}$ is a local vertex.  This factor
is simply the Jacobi identity. Thus the $f^{abc}$'s are the structure
constants of any semisimple group times a product of an arbitrary number of
$\uo$ factors, as might have been expected.  The resulting vertex $V_{4o}$
is the same as that of the heterotic string \cite{OVh} (excluding the
gravitational sector), and gives a correction to the lagrangian of
eq\eq{cubic}.  The modified lagrangian will be further discussed in the next
section.

\subsection{Closed string group theory?}

Since we have taken an agnostic approach to inserting group theory factors
into string diagrams, we can ask if it is also possible to add such
factors in the closed sector.  Thus consider adding an index $a$ to the
closed-string scalar, and take the three-point function on the sphere to be
\beq
A_{ccc} \eqq \kappa \, c_{12}^2 \, d^{\p1 abc} \comma
\eeq
where, in this case, $d^{\p1 abc}$ is a totally symmetric tensor.
The field-theory amplitudes are calculated using the vertex derived from the
action
of (\ref{pleb}), in which $c_{12}^2$ is replaced by \cite{OV}
\beq
\tilde{c}_{12}^2 \sim c_{12}^2 -s^2 \stop
\eeq
(This difference is a field redefinition that affects only local terms.)
The four-point field theory amplitude is given by
\beq
A_{cccc\,\hbox{\scriptsize FT}} =    \kappa^2 \left\{
\f{\tilde{c}_{12}^2 \tilde{c}_{34}^2}s \, d^{\p1 abx} d^{\p1 xcd}
+ \f{\tilde{c}_{23}^2 \tilde{c}_{41}^2}t  \, d^{\p1 cbx} d^{\p1 xad}
+ \f{\tilde{c}_{31}^2 \tilde{c}_{24}^2}u \, d^{\p1 cax} d^{\p1 xbd} \right\}
- V_{4c}\stop
\label{ftc}
\eeq
Using the identities of eq\eq{iden}, this can be rewritten as
\beq
\eqalign{
A_{cccc\,\hbox{\scriptsize FT}} &=    \kappa^2 \left\{
s \, (A^2+s^2-c_{12}^2-c_{34}^2) \, d^{\p1 abx} d^{\p1 xcd} + t \, (B^2
+t^2-c_{23}^2-c_{41}^2) \, d^{\p1 cbx} d^{\p1 xad} \right. \cr
& \left.
\qquad + u \, (C^2 +u^2-c_{31}^2-c_{24}^2) \, d^{\p1 cax} d^{\p1 xbd} \right\}
- V_{4c}\comma \cr
&\propto \vphantom{\nfrac12}
s A^2 \left ( d^{\p1 abx} d^{\p1 xcd} - d^{\p1 acx} d^{\p1 xbd} \right)
+t B^2 \left ( d^{\p1 cbx} d^{\p1 xad} - d^{\p1 cax} d^{\p1 xbd} \right)
+ \hbox{\sl local}
\stop \cr
}
\eeq
In order for this amplitude to be zero, in accordance with the string
amplitude of eq\eq{a4c},
the factors multiplying $s A^2$ and $t B^2$ must separately vanish.
This means that the $d^{\p1 abc}$ factors are essentially trivial:  Define
the matrices
\beq
(D^b )_{ax} \equiv d^{\p1 abx} \stop
\eeq
The vanishing of the factor multiplying $s A^2$ tells us that this set
of matrices is mutually commuting.  They can therefore be
simultaneously diagonalized by an orthogonal matrix $O$, giving
\beq
( D^{\p1 \prime \, b} )_{a'c'} = (O^T D^b O)_{a'c'} = \delta_{a'c'} \, m_{a'}^b
\quad\hbox{(no sum)} \stop
\eeq
Performing this orthogonal rotation on the fields of the theory,
the modified $d^{\p1 \prime}$'s become
\beq
d^{\p1 a'b'c'} = \delta_{a'c'} \, m_{a'}^{b'} \quad\hbox{(no sum)} \comma
\eeq
where $m_{a'}^{b'} = O_b^{b'} m_{a'}^b$.  Since $d^{\p1 a'b'c'}$ is totally
symmetric, this equation means that it must be completely ``diagonal'':
\beq
d^{\p1 a'b'c'} = \delta_{a',b',c'} \, m(a') \stop
\eeq
Thus the theory is simply a sum of noninteracting copies of the theory
without the group factors.  Returning to eq\eq{ftc}, one can now see
that $V_{4c}=0$.  For the rest of this paper, we shall not consider
the rather uninteresting case of a sum of separate theories,
and shall return to the case of a single closed-string scalar, without group
factors in the closed sector.

Note that the $\ZZ_n$ theory of Ooguri and Vafa
\cite{OV} is an example of a closed string with such group-theory
factors.  The theory contains $n$ fields $\phi^l$, which are considered to
have ``spin'' $l/n$ with respect to the $\uo$ part of the \Kahler{}
Lorentz group $\uoo$.  However, while the little group of a massless
scalar with a full $\sott$ Lorentz group would be $\RR$, this group is
completely broken in the \Kahler{} case.  The notion of the spin of a
particle therefore no longer exists, and the theory becomes a theory of $n$
``scalars''.  Since its three-point vertex is of the form $\tilde{c}_{12}^2$
times constant
factors, the theory must be the sum of $n$
noninteracting theories, by our arguments above.  This was indeed found
to be the case in ref.~\cite{OV}, after
taking appropriate linear combinations of the fields.

\subsection{Mixed open/closed amplitudes.}

Thus far, we have collected all the pure open and closed three and
four-point amplitudes.  Since the effective action of the open theory
also contains terms mixing the gauge and gravitational sectors, we shall
also need to consider mixed amplitudes.  We shall thus calculate the
remaining three-point amplitudes, and another four-point amplitude, to
lead us to the full equations of motion of the theory.

Mixed open/closed amplitudes are relatively unfamiliar, so we shall
first review a few basic facts that we shall need:
The Green functions on the upper-half plane ($Im z \ge 0$) are:
\beq
\eqalign{
\VEV{ x^i (z) \xb_j (w) } \phantom{\psi}
&= - \delta_j^i \, \Bigl( \, \log |z-w| + \log|z-\wb| \, \Bigr ) \cr
\VEV{ \psi_R^i (z) \psb_{R j} (w) } &= \delta_j^i \, \f1{z-w} \cr
\VEV{ \psi_L^i (z) \psb_{L j} (w) } &= \delta_j^i \, \f1{\zb - \wb} \cr
\VEV{ \psi_R^i (z) \psb_{L j} (w) } &= \delta_j^i \, \f1{z - \wb} \cr
\VEV{ \psb_R^i (z) \psi_{L j} (w) } &= \delta_j^i \, \f1{z - \wb} \stop
}
\eeq
The UHP has three {\it real\/} bosonic \Mobius{} transformations, and
$2N$ real fermionic ones (with $N$ the number of supersymmetries).
Since closed-string
vertices are parameterized by a {\it complex\/} point in superspace, even
three-point functions with closed vertices involve some integrations over
the positions of the vertices.  If one gauge fixes the fermionic
transformations by demanding that some $\theta$'s be put to zero, the
jacobian of the super-\Mobius{} group splits into a product of bosonic and
fermionic jacobians.  Denoting positions of open vertices by $a,b,c
\ldots$, and those of closed vertices by $z=x+i y, \ldots$, one finds
that the bosonic jacobians for various gauge fixings are
(in an obvious notation):
\beq
\eqalign{
J_{abc} &= |(a-b)(b-c)(c-a)| \cr
J_{az} &= y |a-z|^2 \cr
J_{aby} &= y |(a-b)(2x-b-a)| \cr
J_{z y'} &= 2 y y' |x-x'| \comma \cr
}
\eeq
while the jacobians of the fermionic transformations are:
\beq
\eqalign{
J_{\theta_a \theta_b} &= 1/|a-b|^N \cr
J_{\theta_z} &= 1/y^N \stop \cr
}
\eeq

We can now proceed with the calculation of the mixed three and four-point
amplitudes, taking them in order of increasing number of closed
vertices.  We first calculate the amplitude of two open strings
with one closed one.
The simplest way to do the calculation is to fix $\theta_a$, $\theta_b$,
$a$ and $z$, and to integrate $b$ over the real line, although of
course the final result is independent of this choice.  One obtains
\beq
\eqalign{
A_{ooc} &= \f\kappa\pi \, \delta^{ab} c_{12}^2 \int_{-\infty}^\infty d b
\, \f{y}{(x-b)^2+y^2} \cr
&= \kappa \, \delta^{ab} c_{12}^2  \stop
\label{aooc}
}
\eeq
Since this amplitude gives the coupling of the gravitational sector to
the kinetic term of the open scalars, we have chosen
to append the group-theory factor $\delta^{ab}$, as in the kinetic
term.   This vertex is the same as the gravitational self-interaction
of eq\eq{a3c}, showing a ``universality'' in the couplings of the
various fields to gravity.  The meaning of the interaction will
be further considered in the next section.

One can also find the amplitude of one closed string with three open
ones, needed for the gravitational coupling of the cubic
Yang-Mills vertex.  A simple gauge-fixing for this amplitude is to
fix the
positions of the three open vertices to be at $0$, $1$ and $\infty$,
and to fix the $\theta$'s of the two vertices at $0$ and $\infty$.  One
then needs to integrate the closed vertex over
the UHP.  Since the integrand is symmetric under
$z \to \zb$, one can transform this integral to an integral over the
entire plane, and use the integration results of
ref.~\cite{openclosed}.  The result is
\beq
A_{oooc} = \nfrac{i}2 \kappa g f^{abc} \, F
         \, \f { \Gamma(s) \, \Gamma(t) \, \Gamma(u) }
                 { \Gamma(-s) \, \Gamma(-t) \, \Gamma(-u) }
                 \, \left( c_{12} t +c_{23} s \right)
\comma
\eeq
which vanishes on shell because of the presence of the $F$ factor of
eq\eq{F}.

We now turn to amplitudes with two closed strings.
Since the open string carries a group index, one would expect that
any amplitude with only one open string and an arbitrary number of closed ones
should vanish.  That this is indeed the case follows from the twist symmetry
of the theory:  The UHP is invariant
under the diffeomorphism $z \to -\zb$.
Since open string vertices are odd under the
diffeomorphism, while closed vertices are even, one has
the relation $\vev{V_o^{1} V_o^{2}\ldots V_o^{n}} = (-)^n
\vev{V_o^{n} V_o^{n-1}\ldots V_o^{1}}$.  This means that
any amplitude with only one open string vanishes.  This conclusion is borne
out for $V_{occ}$ by explicitly examining the integrand of the amplitude,
an exercise which we shall spare the reader.

The above argument shows that we do not need to consider the
four-point amplitude with one open and three closed strings.  The only
remaining four-point amplitude contains two open and two closed
strings.  Since its calculation involves carrying out a complicated
three-dimensional integral, and since it should presumably again vanish
by the presence of an $F$ factor, we shall not perform
the calculation.

A more interesting amplitude, and the last one we shall consider, is
the scattering of three closed strings at genus $\nfrac12$.  Unlike
the previous amplitudes, these diagrams can be thought of as giving
a quantum correction to the three-point amplitude on the sphere, as
given in eq\eq{a3c}.  (We
remind the reader that the notion of a tree-level amplitude in the
open string is not that well defined, and that one has the relation
$\kappa \sim \hbar g^2$ between the closed and open-string couplings.)
The amplitude on the UHP is calculated by fixing $z$ and $\theta_z$ of one
vertex, and $y'$ of a second vertex, but
still involves the integration over $x'$ and the position of the third
vertex.  Examining the integrand, one sees that the answer has
the form
\beq
A_{ccc}' \propto \f{\kappa^3}{g^2} \, c_{12}^4 \comma
\label{half}
\eeq
with some finite coefficient\footnote{Which Mathematica thinks is $-2/3$.}.
Since this amplitude contains no open-string vertices, it has a different
structure to all the previous amplitudes, and needs to be allocated some
(unknown) group-theory factor.  In addition, there is a similar
graph on the projective plane $\rptwo$, which should be combined with
the UHP graph.  (In the Chan-Paton scheme the resulting
group-theory factor would be $N-2$ for the group $\son$, \etc)  We
shall consider this counterterm with an arbitrary coefficient in the
following, although the most reasonable choice would be to take it to
vanish.

It is interesting that, in the corresponding amplitude on the torus,
one finds a similar result \cite{italians}:
\beq
A_{ccc}'' \sim {\kappa^3}\, c_{12}^6 \comma
\label{one}
\eeq
but with a coefficient that diverges in the infrared!

\subsection{The partition function}

In addition to scattering amplitudes, one would also like to find
the partition function
of the theory.  In an open theory this involves calculating the path
integral on all the genus zero graphs: the torus, the Klein bottle, the
\Mobius{} band and the
annulus.  The torus amplitude has been calculated in ref.~\cite{part}, and
was found, in their normalization, to be the volume form on moduli space:
\beq
Z_{\torus} = \f1{4\pi} \int_{\cal M} \f{d \tau d \bar\tau}{\tau_2^2} \p6
\left( = \f1{12} \right) \stop
\label{torus}
\eeq
Note that, because of the extra ghosts of the $\uo$ symmetry, the
partition function appears as the partition function of a scalar in 2
{\it real\/} dimensions.  In particular, this means that
if one compactifies two of the
dimensions, as was done in ref.~\cite{OV} using the techniques of
ref.~\cite{kapl}, one again finds an infrared divergence.

Instead of performing the path-integral for our case, we shall obtain the
partition function from the
free energy \cite{Polchinski}, knowing
the spectrum of the open and closed sectors of the theory (considered
to live in two dimensions).
Recall that, while the torus amplitude is
defined over the keyhole region of moduli space, the moduli space of
the other surfaces is simply the real line, and their partition
functions are integrated over the same range as their free energies
(see, for example, ref.~\cite{menew}).
Defining the proper time $t$, the total partition function is given by
\beq
Z = \nfrac12 Z_{\torus} +
\f1{4\pi} \int_0^\infty \f{d t}{t^2} \left(\, \nfrac12 + \nfrac12 \,
c_{\annulus}
+ \nfrac12 \, c_{\mobius} \right)
\label{partfn}
\comma
\eeq
where the coefficients $c$ are group-theory factors.  The
factors of $1/2$ are due to the nonorientability of the theory, and
should be dropped if nonorientable graphs are discarded.  We have
put $c_{\klein} = 1$, since we want the sum of the contributions of the
torus and
Klein bottle to be that of the one scalar of the closed sector.  In the
Chan-Paton scheme one would have
$c_{\annulus} = N^2$, and $c_{\mobius} = \pm N$ for
the groups $\son$ and $\spn$, resulting in a correct description of
the spectrum of the open sector.  If one chooses to implement a more
general \ansatz{} for the group $G$, one needs to enforce the
condition
\beq
c_{\annulus} + c_{\mobius} = 2 \dim{G} \comma
\eeq
which is rather unnatural.

Note also that the partition function of eq\eq{partfn} shows yet
another infrared divergence as $t \to 0$.  In the $N=0$ and $N=1$
theories, such divergences can be isolated by
performing
modular-like
transformations rotating the diagrams by $90^\circ$.
We thus define $q$ for the three graphs by \cite{menew}:
\beq
\eqalign{
q_{\klein}  &\equiv e^{-\pi/t} \cr
q_{\annulus}  &\equiv e^{-4\pi/t} \cr
q_{\mobius}  &\equiv - e^{-\pi/t} \stop \cr
}
\eeq
The partition function then becomes
\beq
Z = \nfrac12 Z_{\torus} +
\f1{16\pi} \left[ \, \int_0^1 \f{d q}{q} \left(2 + \nfrac12 \, c_{\annulus}
\right) +
\int_{-1}^0 \f{d q}{q} \, 2 c_{\mobius} \, \right] \stop \label{part}
\eeq
Thus, in the Chan-Paton case, the divergence at $q=0$ can be regulated by
a principle-part prescription for the (very uninteresting) group
$\sot$.  This is the analogue of the groups $SO(32)$ for the
superstring \cite{famous} and $SO(8192)$ for the bosonic string \cite{menew}.
However, since there are no known anomalies for the bosonic and $N=2$
strings, the argument for these special groups is not very compelling in
these cases,
especially since the theories retain similar
divergences, even when these special groups are chosen.

\section{Space-time interpretation of the open string.}

In the previous section, we began constructing actions for the purely open
and purely closed sectors. Here we shall continue this process, using the
three-point amplitudes we have found
and the vanishing of the four-point amplitudes to
determine the effective field theory string up terms quartic in the fields.
This will be sufficient to allow us to deduce the complete equations of
motion of the theory.

The actions of the open and closed sectors to cubic order are given in
eqs\eq{pleb} and (\ref{cubic}), using the open and closed three-point
amplitudes of eqs\eq{a3c} and (\ref{a3o}). Similarly, the three-point
vertex of one ``graviton'' and two ``gluons'', given in eq\eq{aooc}, can be
obtained from the interaction
\beq
{\cal L}_{ooc} = \int \dxf \, \left( 2\kappa \, \phi \, \pa\pb \varphi^a \wedge
\pa\pb \varphi^a \right) \stop
\label{looc}
\eeq
This gives the complete action to cubic level, excluding the quantum
``counterterms'' of eqs\eq{half} and (\ref{one}) which we shall discuss
later.

The quartic terms in the action are found by calculating four-point
field-theory amplitudes, and demanding that they vanish.  In the discussion
following eq\eq{ftc} (without closed group factors), we saw that
$V_{4c}=0$, showing that the closed action does not receive quartic corrections
\cite{OV}.  In the open sector, on the other hand, we saw that one needs a
nonvanishing vertex $V_{4o}$ to cancel the result of the graphs built from
two $A_{ooo}$'s.  $V_{4o}$ is determined
from eq\eq{V4}, and can be derived from the interaction
\beq
{\cal L}_{4o} = \int \dxf \, \left(- \f{g^2}6 \, f^{adx} f^{xbc} \,
\pa^i \varphi^a
\varphi^b \pb_{\ib} \varphi^c \varphi^d \right) \comma
\label{loooo}
\eeq
in agreement with the result of the pure Yang-Mills sector of the
heterotic string \cite{OVh}.  Note that, unlike the case of the heterotic
string, one does not receive additional contribution from graphs with an
intermediate graviton.  In an open theory such graphs have the topology
of an annulus, and should be considered to be true quantum corrections to the
theory.

There is also a mixed quartic term in the action, which
is found by examining the four-point function
with one closed and three open scalars.  In the field theory, the new
vertex $V_{oooc}$ must cancel the graphs made by sewing together the vertex
$A_{ooo}$ (from eq\eq{a3o}) with the vertex $\tilde{A}_{ooc} = \kappa
\, \delta^{ab} \,
\tilde{c}_{12}^2$ derived from the interaction of eq\eq{looc}.
Calling the closed scalar particle 1, and summing over the three channels,
one obtains
\beq
\eqalign{
A_{oooc\,\hbox{\scriptsize FT}} &=
-i \kappa g f^{abc} \left\{
\left(\f{c_{12}^2}s-s \right) c_{34} -
\left(\f{c_{13}^2}u-u \right) c_{24} +
\left(\f{c_{14}^2}t-t \right) c_{23}
\right\} -V_{oooc}  \cr
&= \vphantom{\nfrac12}
-i \kappa g f^{abc} \left\{
A \, c_{12} -s \, c_{34} + C \, c_{13} +u \, c_{24}-B \, c_{14} -t \, c_{23}
\right\} -V_{oooc}  \cr
&= \vphantom{\nfrac12}
-2i \kappa g f^{abc} \left\{
t \, c_{13} - u \, c_{14}
\right\} -V_{oooc}  \stop \cr
}
\eeq
The resulting vertex can be derived from the interaction
\beq
{\cal L}_{oooc} = \int \dxf \, \left( - \nfrac43 \, i g \kappa \, f^{abc} \,
\pa \pb \phi \wedge \varphi^a
\pa \varphi^b \pb \varphi^c \right) \stop
\label{loooc}
\eeq

Finally, we can consider the amplitude with two closed and two open
scalars.  This can be obtained in two ways: by sewing together two
$\tilde{A}_{ooc}$'s with an intermediate gluon, or by sewing an $A_{ccc}$
to an $\tilde{A}_{ooc}$ with an intermediate graviton. Since these vertices
are both essentially $\tilde{c}^2$, the result (including relative symmetry
factors) is proportional to the amplitude of four closed strings, which we
know to vanish.  There is thus no term coupling two gravitons to the open
lagrangian.  Redefining $\phi$ by a factor of $4\kappa$, and $\varphi$ by a
factor of $2g$, and taking $\varphi$ to be an antihermitian matrix in the
Lie algebra, the total lagrangian to this order %of ${\cal L}_{3o}$, ${\cal
%L}_c$, ${\cal L}_{ooc}$, ${\cal L}_{4o}$ and ${\cal L}_{oooc}$
can be written as
\beq
\eqalign{
{\cal L} &= \f1{16\kappa^2} \, \int \dxf \, \left(
\nfrac12 \, \pa^i \phi \, \pb_{\ib} \phi +
\nfrac16 \, \phi \, \pa \pb \phi \wedge \pa \pb \phi \right) \, \cr
& +
\f1{4g^2} \, \int \dxf \, \left(
-\nfrac12 \, \Tr\Bigl(\pa^i \varphi \pb_{\ib} \varphi \Bigr)
- \f{i}{3!} \, \Tr\Bigl(\pb_{\ib} \varphi \left[\pa^i \varphi, \varphi
\right]\Bigr)
+ \f{1}{4!} \, \Tr\Bigl(\pb_{\ib} \varphi \left[\left[\pa^i \varphi, \varphi
\right], \varphi \right]\Bigr) + \cdots \right. \cr
& \phantom{+\f1{4g^2} \, \int \dxf \, \quad}
\left. + \nfrac12 \, \pa \pb \phi \wedge
\Tr\Bigl(\varphi \pa \pb \varphi \Bigr)
+ \f{i}{3!} \, \pa \pb \phi \wedge
\Tr\Bigl(\varphi \left[\pa \varphi, \pb \varphi \right]\Bigr) + \cdots
\right) \stop
\label{L}
}
\eeq

This lagrangian appears to be rather arbitrary, but its resulting
equations of motion can be given a simple geometrical interpretation if
one considers the
\twot{} dimensional space to be intrinsically \Kahler{} \cite{OV}.
One can then define the \Kahler{} potential
\beq
K = \eta_{i \jb} x^i \xb^{\jb} + \phi \comma
\label{kahler}\eeq
giving rise to the spacetime metric $g_{i \jb} = \pa_i \, \pb_{\jb} \, K=
\eta_{i \jb} + \pa_i \, \pb_{\jb} \, \phi$.
Consider
first the purely closed terms in ${\cal L}$.  With this interpretation, the
equation of motion of $\phi$ becomes the Plebanski equation
\cite{Plebanski}
\beq
\det{g_{i \jb}} = -1 \stop
\label{pleb3}
\eeq
This equation can also be written in terms of forms as
\beq
\pa \pb K \wedge \pa \pb K = 2 \omega \wedge \bar\omega \comma
\eeq
where $\omega$ is the holomorphic $(2,0)$ form $dx^1 \wedge dx^2$ on the
space. As discussed in ref.~\cite{OV}, the Plebanski equation leads to the
vanishing of the Ricci tensor of the \twot{} space, and is equivalent to
the statement that the space have a self-dual Riemann tensor.  The closed
theory is thus a theory of self-dual gravity.

The equation of motion in the purely open sector also has a simple meaning in
(flat) \Kahler{} space \cite{OVh}---it is Yang's equation for SDYM
\cite{Yang}.  This can be seen by defining:
\beq
\eqalign{
A &\equiv e^{-i \varphi/2} \pa e^{i \varphi/2} \; \implies \; F
\wedge \bar\omega = 0\cr
\bar{A} &\equiv e^{i \varphi/2} \pb e^{-i \varphi/2} \; \implies \; F \wedge
\omega = 0\stop
\label{yang}
}
\eeq
The equation of motion of $\varphi$, to this order and ignoring $\phi$, is
\beq
\pb_{\bar{i}} \left( e^{-i \varphi} \pa^i e^{i \varphi} \right) = 0
\; \implies \; F \wedge k_0 = 0
 \comma
\label{flat}
\eeq
where $k_0$ is the flat \Kahler{} form.  Eqs\eq{yang} and (\ref{flat}) are
equivalent to the self-duality of the Yang-Mills field strength in the
\Kahler{} space.

We can now easily find the interpretation of the equations of motion of the
full lagrangian ${\cal L}$ of eq\eq{L}.  The equation of motion of $\varphi$
is the obvious generalization of eq\eq{flat}\footnote{In this equation the
indices are raised with the $\epsilon$ symbol.}:
\beq
g_{i \jb} \, \pb^{\bar{j}} \left( e^{-i \varphi} \pa^i e^{i \varphi} \right) =
0
\; \implies \; F \wedge k = 0     \comma
\eeq
where $k$ is now the full \Kahler{} form of the \Kahler{} potential of
eq\eq{kahler}.  With the definitions of eq\eq{yang}, this is the statement
of the self-duality of the Yang-Mills field strength on
the curved
\Kahler{} space.  On the other hand, the Ricci-flatness condition
of the closed theory gets modified.  The full equation of motion
of $\phi$ is:
\beq
\pa \pb K \wedge \pa \pb K = 2 \omega \wedge \bar\omega
- \f{4\kappa^2}{g^2} \, \Tr \, \left( F \wedge F \right)
\comma
\label{sd1}
\eeq
or
\beq
\det{g_{i \jb}} = -1 - \f{2\kappa^2}{g^2} \, \Tr \left( F_{i \jb} F^{i \jb}
 \right)\stop
\label{sd2}
\eeq
One thus sees that, as in Einstein gravity, the matter sector provides a
source term to the gravitational field equations.  If one was to include
the gravitational counterterms of eqs\eq{half} and (\ref{one}), coming from
the closed scattering amplitudes on the higher-genus diagrams, the
right-hand side of eq\eq{sd1} would also get corrections
of the form $R \wedge R$, and terms with higher derivatives.

Since all these equations have a clear meaning, it is reasonable to assume
that they are, in fact, the equations of the string theory to all orders in
the fields.

\section{Conclusions}

As in the closed and heterotic \nq{} theories, the
\nq{} open string theory in \twot{} dimensional spacetime has
the peculiar property that its four-point scattering amplitudes vanish
identically.  All these theories can be written as ``scalar'' field
theories in a two-complex dimensional spacetime, although the heterotic
string has to be dimensionally reduced.  The closed theory has the
interpretation of being a theory of self-dual gravity \cite{OV}.  The
heterotic string is related to a theory of SDYM, but not in a
straightforward manner \cite{OVh}.
First, the theory must be reduced to a $(2,1)$
dimensional spacetime, in which the particles become tachyonic, and second,
the Yang-Mills self-duality equation receives corrections from diagrams
with internal gravitons. By contrast, we have seen that the open theory is
simply described by SDYM in the \Kahler{} background of the closed sector.
The gravitational equation of motion is however modified by terms of the
form $\Tr \, \left( F \wedge F \right)$, so the resulting spacetime is no
longer self-dual.

Because of the ``topological'' nature of the \nq{} theories, with all
amplitudes being local, the usual constraints on string theories coming
from unitarity and factorization are greatly weakened (assuming that we
understand these issues in spacetimes with two time-like directions). In
particular, the open theory can be defined for any gauge group, and one
does not have to use the standard Chan-Paton \ansatz{}.  At this stage,
there is little to constrain more general \ansatze{}, although their
\adhoc{} nature may be considered rather unpleasant. The further
calculation of loop amplitudes in all \nq{} strings may be important for
clarifying their nature, and, in particular, knowing loop amplitudes of
\nq{} open strings may further constrain the allowed SDYM groups.  Infrared
divergences in the partition function suggests that in the standard
Chan-Paton \ansatz{}, the trivial group $\sot$ may be favoured, analogously
to the group $SO(32)$ of the superstring, but this conclusion is very weak.

In general, our understanding of \nq{} strings is still in its infancy. The
theories appear to be related to self-duality equations in \twot{}
dimensional spaces, but the spacetimes have to be taken as being basically
\Kahler, and the expected Lorentz invariance is absent.  Amplitudes in the
theories are local, or vanish identically, but this is understood only from
explicit calculation. Calculations of loop amplitudes also indicate
peculiar problems.  One-loop results show basic infra-red divergences, and
the string theory results appear to agree with those of the effective field
theories only if the theories are considered as being in two,
and not in four dimensions.
Clearly, there should be deep results to be discovered
here.

\vskip 1 cm

{\bf \noindent Note added.}

While this paper was in preparation, we received two papers from W.~Siegel
\cite{Warren}.  By considering the Ramond sector, one of the continuously
degenerate sectors of the theory, he argued that the theory should actually be
described by an $N=4$ supersymmetric SDYM theory. He also argued that the
$N=2$ string is the same as the ``$N=4$'' $\sut$ string of
ref.~\cite{adem}, and that the theory should then be able to be
written in a fully
$\sott$ Lorentz-invariant form.

\vskip 1 cm

{\large \bf \noindent Acknowledgments}

I am grateful to Y.~Oz for an enlightening conversation.

\newpage

{
\small
\def\baselinestretch{1}
\parskip=-4pt plus 2pt

}

\end{document}